\documentclass[runningheads]{llncs}
\usepackage{graphicx}
\usepackage{amsmath,amsfonts,amssymb}
\usepackage{graphicx}
\usepackage{algorithm}
\usepackage{algpseudocode}

\begin{document}
\title{The Quantum Version of Prediction for Binary Classification Problem by Ensemble Methods\thanks{A part of the reported study is funded by RFBR according to the research project No.20-37-70080. The research is funded by the subsidy allocated to Kazan Federal University for the state assignment in the sphere of scientific activities, project No. 0671-2020-0065.}}
\author{Kamil  Khadiev\inst{1,2} \and
Liliia Safina\inst{1}}
\authorrunning{K. Khadiev, L. Safina}
\institute{Institute of Computational Mathematics and Information Technologies, Kazan Federal University, Kremlyovskaya, 35, Kazan, Russia\and
Zavoisky Physical-Technical Institute, FRC Kazan Scientific Center of RAS, Kazan, Russia}
\maketitle              % typeset the header of the contribution
\begin{abstract}
In this work, we consider the performance of using a quantum algorithm to predict a result for a binary classification problem if a machine learning model is an ensemble from any simple classifiers. Such an approach is faster than classical prediction and uses quantum and classical computing, but it is based on a probabilistic algorithm. Let $N$ be a number of classifiers from an ensemble model and $O(T)$ be the running time of prediction on one classifier. In classical case, an ensemble model gets answers from each classifier and ``averages'' the result. The running time in classical case is $O\left( N \cdot T \right)$. We propose an algorithm which works in $O\left(\sqrt{N} \cdot T\right)$.

\end{abstract}
\section{Introduction}
\label{sec:intro}  % \label{} allows reference to this section

In this work, we suggest a new approach to predict a result for binary classification problem \cite{ML1} by ensemble methods \cite{zhou2019ensemble,schapire2013boosting}.
The key idea is to speed up prediction using quantum algorithm of amplitude estimation \cite{bhmt2002,dewolf2021quantum} as a subroutine.

In recent decades the quantum computing becomes more popular in machine learning \cite{KSV02,ko2018,aazksw2019part1,aazksw2019part2,kms2019,kms2021,ks2021}. Machine learning algorithms require more computing power because of the large amount of data. Properties of quantum algorithms can be useful to solve that problem. 

The combination of small (or not small) models of machine learning is named ensemble methods. The most famous ensemble models  are random forest and gradient tree boosting \cite{breiman1996stacked}. These models are constructed by small trees \cite{DT,c50main}. Each tree returns a result. After that, an ensemble model "averages" using some approach the final result. We store the result in a quantum state.

A quantum register is presented by summation of all possible states:
\[|x\rangle = \sum_{i=1}^{N}{a_i |x_i\rangle},\]

where $a_i$ is an amplitude of a state $|x_i \rangle$. Squared amplitude $a_i^2$ is a probability to get a state $|x_i \rangle$ after measurement. Our method of using quantum subroutines to speed up a prediction process for binary classification problems is based on quantum amplitude estimation. 

Each classifier returns a number of class, first class or second class. Let $Class_1$ be sign of first class, then $Class_2$ be a sign of second class. Let us to label all $Class_1$ results as $Good$, and $Class_2$ as $Bad$. Therefore a quantum register after running a prediction process becomes:

\[|res\rangle = \sum_{i=1}^{N}{a_i |res_i\rangle} = a_g |Good\rangle + a_b|Bad\rangle,\]
\[a_g =  \sum_{i=1}^{N}{\{a_i |res_i\rangle}: |res_i\rangle \in Good \},\]
\[a_g^2 + a_b^2 = 1.\]

We want to estimate $a_g$.

The structure of this paper is the next. In Section 2 we consider the quantum amplitude amplification algorithm, its running time, and two subroutines to estimate an amplitude of good states and present our amplitude estimation algorithm. In Section 3 we consider the definition of an ensemble method and how to get a final result class for a new input object. 

\section{Amplitude Amplification and Estimation}\label{sec_2}
\subsection{Amplitude Amplification}
Suppose there is a set of good and bad elements. Let $p$ be the probability of finding a good element. Then, in the classical case, we need to repeat random selection process for a set $O\left(\frac{1}{p} \right)$ times to find the good element.

Quantum amplitude amplification algorithm\cite{bhmt2002} allows us to find a good element for $O\left(\frac{1}{\sqrt{p}} \right)$. It is based on Grover's searching algorithm \cite{g96}. Let $A$ be any quantum algorithm that acts to zero state $A |0 \rangle$. Let $\chi(x)$ be a Boolean function, that separates elements for good and bad, $\chi(x) = 1$ if $x \in Good$, and $\chi(x) = 0$ otherwise. The function $\chi$ participates in changing the sign of the amplitude.

It uses the next unitary operator:

\begin{equation}
Q = -A S_0 A^{-1} S_{\chi}
 \label{eq:ref}
\end{equation}

where 

\[S_{\chi} |x \rangle\ =
\begin{cases} 
-|x \rangle\,  \text{if $\chi(x)$=1} \\
|x \rangle\,  \text{otherwise},
\end{cases}
\]

and 

\[S_{0} |x \rangle\ =
\begin{cases} 
-|x \rangle\,  \text{if $|x \rangle=|0 \rangle$} \\
|x \rangle\,  \text{otherwise},
\end{cases}.
\]

\begin{lemma}
The  Quantum Amplitude Amplification works in $O\left(\frac{1}{\sqrt{p}}\right)$.
\end{lemma}

If we know the probability $p$ we can use the amplitude amplification algorithm and find a good element in a set with quadratic speedup. In case when we do not know the value $p$, we can apply amplitude estimation algorithms. Let us consider 4 algorithms to estimate an amplitude of a $Good$ element in a quantum state.

\subsection{Quantum Amplitude Estimation} \label{sec_2_2}
Let us consider  quantum algorithms to estimate amplitude. Two of them are constructed by us.
\subsubsection{Algorithm QSearch}

This subroutine is based on a quantum amplitude amplification algorithm and is created by the authors. It uses a quantum algorithm $A$ that acts on quantum state and $\chi$, it is a Boolean function, that divides on $Good$ and $Bad$ elements. Steps of \textbf{QSearch} is next:

%поменяла некоторые пункты в алгоритме, но я не уверена, что я права. мне кажется, что так логичнее. но надо проверить. 
\begin{enumerate}
    \item Apply $A |0 \rangle$ and measure a system. Let $|z \rangle$ be a measured result. If $\chi(z)$ is $Good$, then stop the algorithm.
    \item Let $l = 0$ and $1 < c < 2$.
    \item $l = l + 1$ and $M = \lceil c^l \rceil$.
    \item Initialize a new quantum register to apply $A$ again. 
    \item Let $j$ be a randomly selected number from $[1, M]$.
    \item Apply $Q^j$ to register, apply the unitary operator $Q$ $j$ times.
    \item Measure the register. Stop if  $\chi(z)$ is $Good$, otherwise go to step 3. 
    \item Return $p =\frac{1}{j}$

\end{enumerate}
\begin{lemma}
\textbf{QSearch} works in $\Theta\left(\frac{1}{\sqrt{p}}\right)$, where $p$ is a probability to find a good element in sequence.
\end{lemma}

\subsubsection{Algorithm Est\_Amp}

Let $F_M$ be the quantum Fourier transform \cite{dewolf2021quantum}:

\[F_M: |x \rangle \rightarrow \frac{1}{\sqrt{M}} \sum_{y= 0}^{M-1} e ^{2 \pi i x y/M}|y \rangle, (0 \leq x < M),\]

where $M\geq 1$ is any integer number.

Let $F_M^{-1}$ be an inverse quantum Fourier transform.

Let $\lambda_M (U)$ be an operator that acts on the unitary operator $U$ as follows:

\[|j \rangle |y \rangle \rightarrow |j \rangle \left(U^j|y \rangle\right), (0\leq j < M),\]

where $M$ is any positive integer value. 

The algorithm also uses the unitary operator $Q$ from Equation (\ref{eq:ref}), a quantum algorithm $A$, a Boolean function $\chi$, and it needs to get an positive integer variable $M$, which affects the accuracy of the count.

Steps of the algorithm \textbf{Est\_Amp $\left(A, \chi, M \right)$} are next:
\begin{enumerate}
    \item Initialize two register
    \item Apply $F_M$ to the first register
    \item Apply $\lambda_M (Q)$
    \item Apply $F_M^{-1}$ to the first register
    \item Measure the first register $|y \rangle $
    \item The output amplitude is $\widetilde{a} = \sin^2{\left(\pi \frac{y}{M}\right)}$
\end{enumerate}
\begin{lemma}
\textbf{Est\_Amp} returns a result $\widetilde{a}$ such that is:

\[|\widetilde{a} - a| \leq 2\pi k \frac{\sqrt{a(1-a)}}{M} + k^2 \frac{\pi^2}{M^2},\]

%я не знаю, что такое к, написать подробнее про к
with probability greater than 0.5, where $k$ is positive integer number.
\end{lemma}
\begin{lemma}
\textbf{Est\_Amp} works in $\Theta\left(\frac{1}{\sqrt{p}}\right)$.
\end{lemma}

Note that $p=\frac{t}{N}$, where $t$ is a size of subset of good elements and $N$ is a size of a given set, $N = |Good| +|Bad|$. In the paper\cite{bhmt2002}
%ссылка 
authors present the methods based on the algorithm \textbf{Est\_Amp} which are found $t$ with high accuracy. 

\subsubsection{Our Algorithm}
In this algorithm we use the quantum amplitude amplification algorithm. It uses $A$ and $\chi$ and has next steps:

\begin{enumerate}
    \item Let $j = 0$, $\epsilon = 0.0001$.
    \item Let $p = \frac{1}{2^j}$.
    \item Apply the quantum amplitude amplification algorithm assuming that the probability of finding a good element is $p$.
    \item If the quantum amplitude amplification algorithm returns a good element ($\chi (x) =1$) or $p < \epsilon$ stop and return a value $p$, or $\sqrt{p}$ if we should know an amplitude of good states.
    \item $j = j+1$. Go to step 2. 
\end{enumerate}
\begin{table}[ht]
\caption{The accuracy of the probability estimation} 
\label{tab:fonts}
\begin{center}       
\begin{tabular}{|l|} %% this creates two columns
%% |l|l| to left justify each column entry
%% |c|c| to center each column entry
%% use of \rule[]{}{} below opens up each row
\hline
\rule[-1ex]{0pt}{3.5ex}  $p \in [l, r) $ or $[l, r]$    \\
\hline
\rule[-1ex]{0pt}{4ex}  $[\frac{1}{2}, 1]$ \\
\hline
\rule[-1ex]{0pt}{4ex}  $[\frac{1}{4}, \frac{1}{2}\big)$ \\

\hline
\rule[-1ex]{0pt}{4ex}  $[\frac{1}{8}, \frac{1}{4}\big)$ \\
\hline
\rule[-1ex]{0pt}{4ex}  $[\frac{1}{16}, \frac{1}{8}\big)$\\
\hline
\rule[-1ex]{0pt}{4ex}  $[\frac{1}{32}, \frac{1}{16}\big)$\\
\hline
\rule[-1ex]{0pt}{4ex}  $[\frac{1}{64}, \frac{1}{32}\big)$\\
\hline
\rule[-1ex]{0pt}{4ex}  $[\frac{1}{128}, \frac{1}{64}\big)$ \\
\hline
\rule[-1ex]{0pt}{4ex} and further  \\
\hline
\end{tabular}
\end{center}
\end{table} 
\begin{lemma}
The running time of our quantum amplitude estimation algorithm is 

\[O\left(\sum_{i=1}^{j}\sqrt{2^i}  \right) = O\left(\sqrt{2^j} \right),\]

where $j \leq 14$ for $\epsilon = 0.0001$.
\end{lemma}

We set such $\epsilon$ because it allows us to estimate an amplitude that can be very small if the probability of finding a good element is almost 0.

If the amplitude of the good element is more than $0.5$, our algorithm can't estimate an amplitude with more accuracy. It only says that an amplitude is more than $0.5$. This point is not critical for problems in which we want to use the algorithm.

\subsubsection{Our Algorithm Using Binary Search}

In classical  case using probabilistic algorithm we can find a good element in $O \left(\frac{1}{p} \right)$, where $p$ is a probability of a good element. Let us consider the next situation: set $p = 0.25$, in this case we should make $\frac{1}{p} = 4$ steps to find a good element. Note that we do not know $p$, but if we find a good element with fixed $p$, then in real $p \geq 0.25$, and $p < 0.25$ otherwise.  
We suggest the next algorithm to estimate an amplitude of a good element. 
It uses $A$ and $\chi$, and it is based on quantum amplitude amplification and binary search \cite{cormen2001} algorithms. 

\begin{enumerate}
    \item Let $left = 0$, $right = 1$, $p = \frac{left + right}{2}$.
    
     \item Apply the quantum amplitude amplification algorithm assuming that the probability of finding a good element is $p$.
     
    \item If the quantum amplitude amplification algorithm returns a good element ($\chi (x) =1$), stop the algorithm and return $p \in [\frac{1}{2}; 1]$; else: $right = p$.
    
    \item $p = \frac{left + right}{2}$.
    
    \item Apply the quantum amplitude amplification algorithm assuming that the probability of finding a good element is $p$.

    \item If the quantum amplitude amplification algorithm returns a good element ($\chi (x) =1$), $left = p$; else: $right = p$.
    \item if $\sqrt{\frac{1}{left }} = \sqrt{\frac{1}{right}}$, then stop and return $p \in [left; right)$. Otherwise go to the step 4. 
    
\end{enumerate}

If $p \geq \frac{1}{2}$ we can say only that $p \in [\frac{1}{2}; 1]$, we can not increase the accuracy. This case is being checked in steps 1-3. After that, we run the binary search approach. It is stopped if the count of steps of left and right borders of a semi-cut are the same. 

The accuracy of the estimation of the amplitude of a good state depends on the value of the amplitude. The accuracy is higher if the amplitude is small. We estimate the segment that the probability belongs to. The probability $p$ can belong to the next segments. Let $j = \frac{1}{p}$ be a number of iteration in the probabilistic algorithm and $\lceil\sqrt{j}\rceil$ --- in the quantum amplitude amplification algorithm.

\begin{table}[ht]
\caption{The accuracy of the probability estimation using binary search approach} 
\label{tab:tabbs}
\begin{center}       
\begin{tabular}{|l|l|} %% this creates two columns
%% |l|l| to left justify each column entry
%% |c|c| to center each column entry
%% use of \rule[]{}{} below opens up each row
\hline
\rule[-1ex]{0pt}{3.5ex}  $p \in [l, r) $ or $[l, r]$ & A number of iterations --- $\sqrt{j}$   \\
\hline
\rule[-1ex]{0pt}{3.5ex}  [0.5, 1]  & $j \leq 2$\\
\hline
\rule[-1ex]{0pt}{3.5ex}  [0.25, 0.5) & $2 < j \leq 4$\\
\hline
\rule[-1ex]{0pt}{3.5ex}  [0.125, 0.25) & $4 < j \leq 8$\\
\hline
\rule[-1ex]{0pt}{3.5ex}  [0.0625, 0.125) & $8 < j \leq 16$   \\
\hline
\rule[-1ex]{0pt}{3.5ex}  [0.046875, 0.0625) & $16 < j \leq 22$ \\
\hline
\rule[-1ex]{0pt}{3.5ex}   [0.03125, 0.046875) & $21 < j \leq 32$\\
\hline
\rule[-1ex]{0pt}{3.5ex}   [0.0234375, 0.003125) & $32 < j \leq 42$\\
\hline
\rule[-1ex]{0pt}{3.5ex}   [0.015625, 0.0234375) & $42 < j \leq 64$\\
\hline
\rule[-1ex]{0pt}{3.5ex} and further & \\
\hline
\end{tabular}
\end{center}
\end{table} 

We can continue to get smaller segments, where $p$ will be less than 0.01, but for our problem is not useful. 
We can not more separate segments in the upper rows of the table because square roots from numbers of iterations will be the same, and we get maximum possible accuracy using such approach. 

In the Table \ref{tab:tabbs} we present the list with segments of $p$. We can say that the number of binary search steps is small. The number of steps in the quantum amplitude amplification subroutine reaches the maximum value with the lowest probability of a good element, and it is about $\sqrt{j} \leq \sqrt{128} \approx 12$ for $p < 0.01$. Hence running time of the algorithm is $O\left(\log 100 \cdot \sqrt{j} \right)$, where the first probability for binary search is $p_{middle}=\frac{l + r}{100}$, $l =0$, $r=100$. This running time is less than $O\left(\sqrt{N}\right)$ for big $N$.  

\section{Ensemble methods in Machine Learning}
In this work, we consider a binary classification problem. Examples of binary classification problems are deciding to give a loan to a bank customer or not, determining the gender of an app user, a sick person or a healthy person, and others. We can reduce a multi-class problem to a binary classification as follows: the classifier $i$ determine whether an object $x$ belongs to the class $i$. In this case, we will have to build $M$ classifiers, where $M$ is the number of classes in the problem.

Each machine learning method has its distinctive properties. The task of the data scientist is to choose the most suitable model for the task. The chosen model should work well not only on the test sample but also be useful for real data. A common problem in the process of training a model is overfitting or underfitting. For example, decision trees are easily overfitted if we construct very deep trees. They work well on training data, but they don't perform well on real data. This problem can be solved by ensemble methods. 

The idea of the ensemble method is to fit some simple and small classifiers, to get results by all of them and to ``average`` a final result. 

The most famous ensemble methods are random forest and gradient tree boosting. Random forest uses deep overfitted trees. Gradient tree boosting, on the contrary, uses small underfitted trees: each subsequent tree clarifies the answer. Moreover, any machine learning models can be used as a small classifier. It can be a decision tree, KNN, K-means, or neural networks, for example, and other models. 

Any small classifiers create a metamodel (ensemble model). It is constructed using bootstrap aggregation, boosting, or stacking approaches. 

The prediction process is the following: each small classifier returns some class, and after voting, a metamodel says a class number for input object $X$. To ``average`` a final result, a metamodel uses the following formula for classification problem: 

\[answer=\max_{c=1}^{M} \sum_{i=1}^{N} f(T_i(X), c)),\]

where

\[
f(r, c) =
\begin{cases} 
1,  \text{if r=c} \\
0,  \text{otherwise},
\end{cases}
\]

where $M$ is a number of classes.

The formula is actual if we assume that each classifier has an equal weight in the metamodel. If we want to determine that some classifier is more accurate than others, we can set weights for each small classifier.

For binary classification, a number of classes is equal to two. Let $N$ be a number of small classifiers. Let $k_1$ be a number of classifiers that returns the first class in the prediction process, and $k_2$ be a number of classifiers that returns the second class. 

\[k_1 + k_2 = N\].

\[
answer =
\begin{cases} 
Class_1,  \text{if $k_1 \geq k_2$} \\
Class_2,  \text{otherwise},
\end{cases}
\].

In other words, we can say that a final result from the ensemble model is $Class_1$ with probability equal to $\frac{k_1}{N}$.

Let $O(T)$ be a running time for prediction on one small classifier from metamodel. In fact, each classifier has its own running time for a prediction. Some models predict faster than others. But let us set the running time of prediction on one classifier by $O(T)$. 

\begin{lemma}
Running time of prediction a result by ensemble method for classification problem is equal to $O(T \cdot N)$. 
\end{lemma}

\section{Probabilistic Algorithm for Prediction}
Let us consider the following probabilistic algorithm to predict a class number for a binary classification problem.
Let $N$ be a number of trained machine learning models.
Let $p_i$ be a probability that the $i$-th model returns the first class. We randomly choose some model with equal probability. The probability of choosing the $i$-th model is $\frac{1}{N}$. Then the probability of getting a result class equal to $Class_1$ is 
\[p = \frac{1}{N} \sum_{i=1}^N p_i\].

Let $model$ be a list of trained machine learning models, $predict(X)$ be a prediction function, it returns a probability of $X$ belongs to the first class, $Get\_Random\_Model(1, N)$ be a function that equally probable chooses a number from 1 to $N$. Steps of probabilistic algorithm is next
\begin{enumerate}
    \item $i = Get\_Random\_Model(1, N)$
    \item Predict a result for input object $X$ on $i$-th model: $p_i = model[i].predict(X)$.
    \item Check $p_i$
    \[
answer =
\begin{cases} 
Class_1,  \text{if $p_i \geq 0.5$} \\
Class_2,  \text{otherwise} 
\end{cases}
\].
\end{enumerate}

Such  approach is not useful for a real problem, but it will be applied in the quantum version of prediction. 
\section{Quantum Amplitude Estimation to Predict a Result Class}

In our quantum version of prediction we use the idea of the probabilistic algorithm above. Let the state $|0 \rangle $ be a $Good$ element ($Class_1$) and $|1 \rangle $ be a $Bad$ element ($Class_2$). Each model returns $Class_1$ and $Class_2$ with some probabilities. In quantum case the prediction process is changing amplitudes process of states $|0 \rangle $ and $|1 \rangle $.

The summary state for $N$ models is:

\[|x\rangle = \frac{1}{\sqrt{N}}\sum_{i=1}^{N}{a_i |x_i\rangle} = \frac{1}{\sqrt{N}}\sum_{i=1}^{N}{\left(g_i |0\rangle + b_i |1\rangle \right)},\]

where $g_i$ is an amplitude of the first class received by $i$-th tree, and $b_i$ is an amplitude of the second class, $\alpha_i^2 = g_i^2 + b_i^2 = 1$. 

Let $p_i$ be a probability of the first class that was returned by $i$-th tree, $p_i = g_i^2$. The common amplitude of $Class_1$ in the quantum state is:

\[a_g = \frac{1}{\sqrt{N}}\sum_{i=1}^{N}{g_i}. \]

Then the total probability of the first class is $p = a_g^2$.

We suggest to compute $p$ using amplitude estimation algorithms. Let $A$ be an algorithm witch predicts $p_i$ for input object $X$, it changes amplitudes of quantum state for $|0 \rangle$ and $|1\rangle$. Let  $\chi (X)$ be a Boolean function that returns $true$ for state $|0\rangle$ ($Class_1$).  Steps of prediction are next:

\textbf{Prediction $\left(models, A, \chi, X \right)$}
\begin{enumerate}
   
    \item $a_g = Amplitude\_Estimation(A, \chi)$
    \item Check $p = a_g^2$
    \[
answer =
\begin{cases} 
Class_1,  \text{if $p \geq 0.5$} \\
Class_2,  \text{otherwise} 
\end{cases}
\].
    \item return $answer$.
\end{enumerate}

The \textbf{Amplitude\_Estimation} is one of subroutines which we described in Section \ref{sec_2_2}.
\section{Conclusion}

In this work, we considered the approach to predict a result class for binary classification problems using classical and quantum algorithms by ensemble methods. The features of quantum computing allow us to parallelize a prediction process on all used machine learning models of the ensemble. We only need to calculate an amplitude of a good quantum state. We use popular quantum amplitude amplification and estimation algorithms and present two our methods based on the quantum amplitude amplification algorithm to estimate an amplitude. 

 \bibliographystyle{splncs04}
 \bibliography{report}
\end{document}